\documentclass[conference]{IEEEtran}
\IEEEoverridecommandlockouts

\usepackage{cite}
\usepackage{amsmath,amssymb,amsfonts}
\usepackage{algorithmic}
\usepackage{graphicx}
\usepackage{textcomp}
\usepackage{xcolor}
\usepackage{soul}
\usepackage{comment}
\usepackage[caption=false,font=footnotesize]{subfig}
\usepackage{siunitx}

\def\BibTeX{{\rm B\kern-.05em{\sc i\kern-.025em b}\kern-.08em
    T\kern-.1667em\lower.7ex\hbox{E}\kern-.125emX}}
\begin{document}

\title{Geometry-Aware LoRaWAN Gateway Placement in Dense Urban Cities Using Digital Twins\\
{\footnotesize \textsuperscript{*}Note: Sub-titles are not captured for https://ieeexplore.ieee.org  and
should not be used}
\thanks{Identify applicable funding agency here. If none, delete this.}
}

\author{\IEEEauthorblockN{1\textsuperscript{st} Abdikarim Mohamed Ibrahim}
\IEEEauthorblockA{\textit{Faculty of Engineering and Technology} \\
\textit{Sunway University}\\
Petaling Jaya, Malaysia \\
abdikarimi@sunway.edu.my}
\and
\IEEEauthorblockN{2\textsuperscript{nd} Rosdiadee Nordin}
\IEEEauthorblockA{\textit{Future Cities Research Institute} \\
\textit{Sunway University}\\
Petaling Jaya, Malaysia  \\
rosdiadeen@sunway.edu.my}
}

\maketitle

\begin{abstract}
 LoRaWAN deployments rely on rough range estimates or simplified propagation models to decide where to place/mount gateways. As a result, operators have limited visibility into how rooftop choice, streets, and building shadowing jointly affect coverage and reliability. This paper addresses the problem of gateway placement in dense urban environments by combining a geometry accurate Digital Twin (DT) with a GPU-accelerated ray tracing engine. Existing studies optimize placement on abstract grids or tune models with sparse measurements; few works evaluate LoRaWAN gateways on a full 3D city model using a realistic link budget. In this paper, we develop a DT with ITU radio materials and evaluate eight candidate rooftops for RAK7289 WisGate Edge Pro gateways under a sub-GHz link budget derived from the data sheet. For each rooftop, we obtain Signal-to-Noise Ratios (SNR) on a 5 meter grid, derive robust and edge coverage indicators, and apply a greedy maximum coverage algorithm to rank sites and quantify the benefit of incremental densification. Results show that a single rooftop gateway covers one fifth of the full Sunway twin (i.e., the DT) at a robust SNR threshold, and that six sites still leave large areas of single gateway or out of coverage cells in surrounding residential streets. The findings from this paper shows that DT and ray tracing tools enable network operators to bridge the gap of expensive real-world trials and planning to identify if the planned LoRaWAN gateway is sufficient or additional sites are required. 
\end{abstract}


\begin{figure*}
    \centering
    \includegraphics[width=0.75\linewidth]{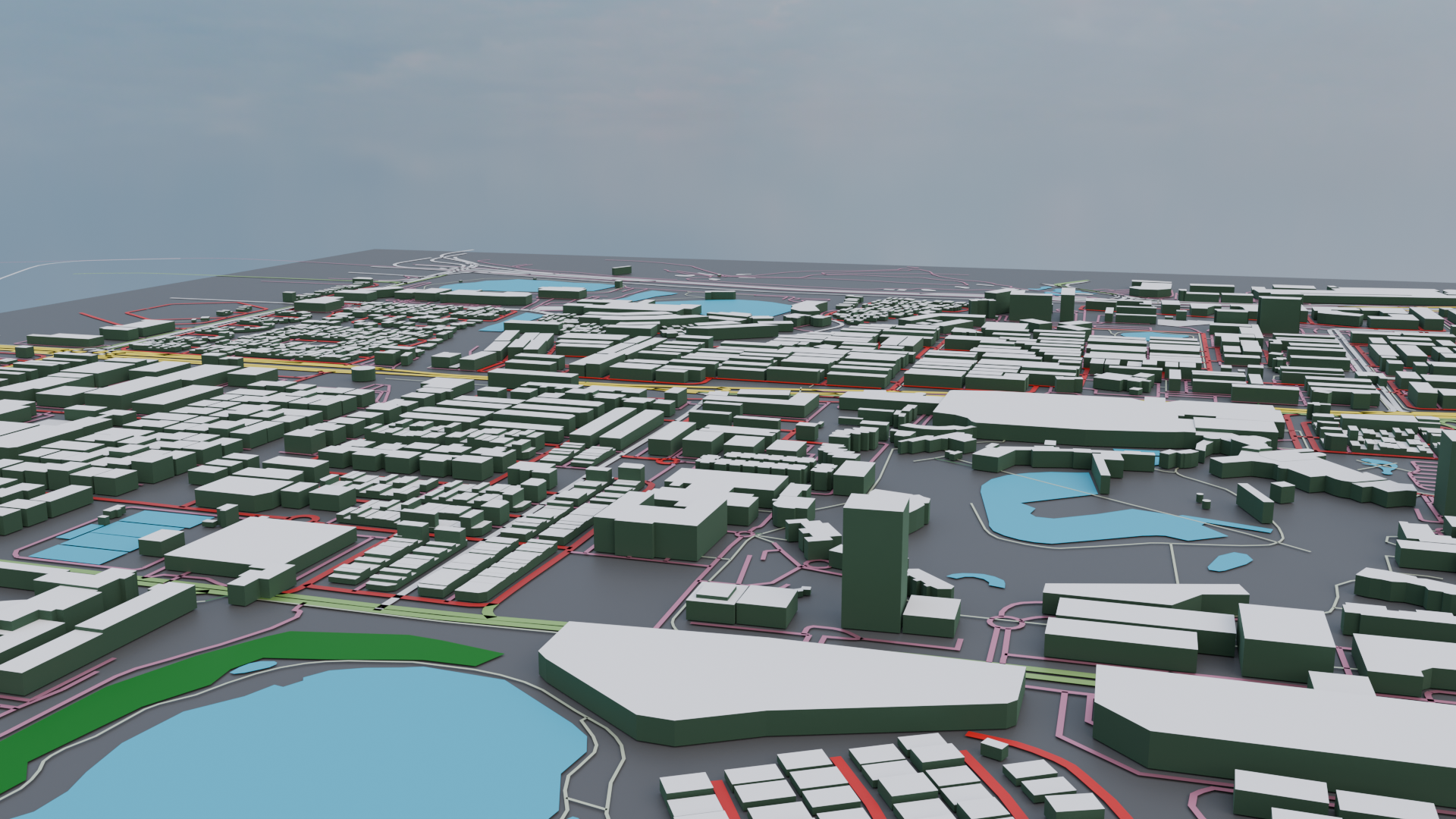}
    \caption{Overview of the Sunway City Twin.}
    \label{fig:sunway_twin_overview}
\end{figure*}

\section{Introduction}
\label{sec:intro}

Low-power wide-area networks (LPWANs) have been one of the main key infrastructure for smart cities as well as rural areas, enabling Internet of Things (IoT) applications such as environmental monitoring \cite{xu2019internet-b2two,tajudin2020blockchain-b1one}, automation \cite{duguma2024internet-b3three}, healthcare \cite{amiri2024applications-b4four}, and connectivity in rural areas \cite{alobaidy2024empowering-b3three3new,abdullah2024indigenous-b4four4new}. Their popularity are attributed to their low power and multi-kilometre coverage with battery lifetimes of many years~\cite{matni2020lorawan-b1Old-now-b6,centenaro2016long-b2old-now-b7}. Among LPWAN options, LoRaWAN is the most used due to its open ecosystem and robust chirp spread spectrum modulation, and it has been shown to perform well in dense urban areas~\cite{villarim2019lora-b8-new}. In these areas, network operators aim for a small number of gateways to support thousands of low-cost sensors. Therefore, the way in which gateways are placed on available rooftops and towers has a direct impact on coverage, reliability, and the cost of the resulting network.

Several studies have investigated LoRaWAN's performance, relying on link-budget analyses, real-world measurement campaigns, and high-level system simulations~\cite{villarim2019lora-b8-new,bor2016lora-b9-new,adelantado2017understanding-b10-new,loh2021robust-b11-new}. However, real-world tests in urban cities have shown mixed results, in which the advertised several kilometers range can rapidly drop in some areas such as tall buildings and street layouts depending on the urban morphology, which makes the exact coverage area irregular. To overcome this issue, studies have developed analytical or empirical path-loss models which have been applied to simplified grids that mimic the actual deployment area to estimate the potential coverage area and the optimal gateway placement \cite{jaeckel2014quadriga-b12now-b6old}. The main issue with these models is the lack of accurate details about the deployment environment. Gateway placement and LoRaWAN performance has also been studied from a network optimization perspective, in which integer programming, theoretical heuristics, and metaheuristics approach have been proposed maximize coverage or minimize the number of gateways~\cite{abdullah2024indigenous-b13-new}. However, most of these studies rely on log-distance path-loss augmented with stochastic shadowing. Therefore, inherently it becomes challenging to capture how the urban morphology of the dense city shape LoRaWAN coverage. More recent work on DTs and GPU accelerated ray tracing has demonstrated that geometry accurate propagation prediction is feasible at city scale~\cite{pegurri2025van3twin-b14-new}, but these tools have so far been applied mainly to higher-frequency 5G and 6G scenarios rather than sub-GHz LPWAN planning.

In this paper we address this gap by combining a geometry accurate DT of Sunway City, Malaysia with a LoRaWAN link budget aligned to a commercial outdoor gateway i.e., (RAK7289V2 WisGate Edge Pro). Sunway City twin as shown in Figure~\ref{fig:sunway_twin_overview}. The twin was designed using Blender with OpenStreetMap and includes building footprints and heights, roads, water bodies, and open spaces over several square kilometres, and is imported into the Sionna ray tracing engine with ITU based radio materials \cite{hoydis2023sionna-b15-new}. In addition to the resulting path gain maps obtained form Sionna we formulate a budgeted maximum coverage problem, in which, we select at most $K$ rooftops from a realistic candidate set so as to maximize the fraction of area that meets specified LoRa Signal-to-Noise Ratio (SNR) thresholds. Specifically, the main contributions of this paper are as follows:
\begin{itemize}
    \item We develop a LoRaWAN planning framework that builds on a DT with ray tracing and an explicit link budget based on commercial outdoor gatewayn.
    \item We formulate gateway placement as a budgeted maximum coverage problem over a DT and apply a greedy optimization algorithm to rank rooftops and quantify coverage as a function of the number of deployed gateways.
    \item We provide analysis of coverage, redundancy (i.e., the number of gateways per cell above given threshold), and best gateway association regions over DT, showing areas where a single gateway deployment is sufficient or not.
\end{itemize}

The remainder of the paper is organized as follows. Section~\ref{sec:related} reviews related work on LoRaWAN coverage modelling, gateway placement, and wireless DTs. Section~\ref{sec:model} presents the Sunway City DT, the LoRaWAN link budget parameters, and the SNR and coverage formulation. Section~\ref{sec:results} presents the numerical results for coverage, redundancy, and association patterns under different gateway budgets. Section~\ref{sec:conclusion} concludes the paper and outlines directions for extending the framework to other LPWAN technologies and deployment scenarios.

\section{Related Work}
\label{sec:related}

Recent studies have explored several aspects of LPWAN to enable massive connectivity including, empirical characterization of LoRaWAN coverage, gateway placement models, and applied intelligence algorithms for wireless planning. Early studies have focused on understanding the basic propagation properties and capacity limits of LoRaWAN. For instance, Centenaro \textit{et al.,} \cite{centenaro2016long-b2old-now-b7} have reviewed LPWAN technologies to answer the question of how LoRa coverage is shaped by deployment morphology. The authors found that LoRa is one of the main enablers of future cities applications. Similarly, Bor \textit{et al.,} \cite{bor2016lora-b9-new} investigated the capacity limits and scalability of LoRa. The authors used experiments to model LoRa link behaviour (i.e., communication range and capture effects), and incorporated the model into a custom simulator. The study showed that LoRa networks scale poorly with static setting and using single sink, but scales better with the use of dynamic communication parameters or increasing the number of sinks. Adelantado \textit{et al.,} \cite{adelantado2017understanding-b10-new} provided an overview of the LoRaWAN technology in terms of capabilities and limitations. The authors examined thees aspects based on use cases and found that that LoRaWAN is not a universal solution for all IoT needs, and its actual capacity is limited by duty cycle regulations. The study recommended that network operates should carefully dimension based on the number of devices and channel parameters to fit the specific use case.

In addition, researchers have also focused on gateway placement as a separate optimization problem using real world trails or propagation models. Loh and Karl formulate robust gateway placement for scalable LoRaWAN as a geometric set-cover problem and use greedy as well as local search heuristics, in order to maximize the number of covered devices with few gateways as possible~\cite{loh2021robust-b11-new}. Khan \emph{et al.,} proposed DPLACE, which is a framework that clusters end devices and then positions gateways to serve these clusters while trading off capital and operational expenditure against coverage and Quality-of-Service (QoS) constraints~\cite{matni2020lorawan-b16-new}. Other studies follow a similar pattern, which involves using log-distance or log-normal path-loss models with simple shadowing terms to approximate urban propagation. These works have been shown to capture the budgeted coverage nature of the placement problem and take advantage of its submodular structure; however, they operate on abstract graphs or regular grids and therefore cannot resolve the impact of urban layout on LoRaWAN coverage.

More recently, DT and GPU accelerated ray tracing engines have been introduced as research tool for wireless planning in  three-dimensional city replicas. Classical deterministic ray tracing have been extensively used in macrocell and microcell planning at microwave and millimetre-wave frequencies~\cite{jaeckel2014quadriga-b12now-b6old}. Modern frameworks such as Sionna and its ray-tracing extension provide differentiable, GPU-accelerated propagation engines that can be embedded to physical-layer simulation~\cite{hoydis2023sionna-b15-new}. Several works have used such tools to study 5G and 6G scenarios, including beamforming, blockage aware cell planning, and vehicular connectivity in dense urban grids~\cite{hoydis2023sionna-b15-new}. However, these studies target centimetre or millimetre wave bands and cellular base stations, not sub-GHz LPWAN gateways. Furthermore, they focus on link level metrics or beamforming strategies, unlike the combinatorial placement of a small number of gateways under several constraints including costs. Therefore, this paper bridges the gap by combining a geometry accurate DT of dense urban city with a LoRa link budget based on a commercial outdoor gateway, and by applying a submodular greedy placement strategy to obtain accurate coverage, redundancy, and association maps for several candidate rooftop gateways.

\section{System Model and LoRaWAN Configuration}
\label{sec:model}

The system model consist of a three-dimensional model of Sunway City, which was developed in Blender with geo-referenced data using OpenStreetMap and exported to Sionna RT. The DT has been designed to match the exact city layout and referenced in a common Cartesian coordinate system in Blender software. Each mesh is defined with a radio material tag that are compatible with the ITU based material library in Sionna RT, so that reflection, diffraction, and penetration losses follow the sub-GHz propagation model implemented in the ray tracing engine. 

The next step involved loading the scene as integrated environment. A regular radio map grid is defined with cell size $(\Delta x,\Delta y)=(5~\text{m},5~\text{m})$ in the horizontal plane. For each transmitter, Sionna computes the complex baseband channel response and the corresponding large-scale path gain towards every grid cell. The ray tracing is then configured with a maximum of three interactions per ray (${max\_depth=3}$) and $2\times 10^6$ rays per transmitter. This is sufficient to stabilize the coverage statistics and keep the computational load manageable.

Secondly, Eight candidate LoRaWAN gateways are placed on the rooftop locations across the scene center (i.e., the university campus). Their $(x,y,z)$ coordinates and heights were taken from the DT and correspond to buildings where an outdoor gateway can be installed. These locations were selected as follows: a) we first manually identified buildings near the university campus in the DT; and b) maintained only buildings that that are high to provide line-of-sight or shallow non-line-of-sight paths towards the campus area, are accessible for power and backhaul, and are not affected or dominated by a neighboring rooftop of similar height and position. This selection criteria led to a set of eight distinct rooftops that cover the main areas of the campus as well as neighboring residential area. 

Each candidate site is modeled as a single element isotropic transmitter with one antenna element and cross polarized pattern ISO in Sionna, which is a representative of sub-GHz omni-directional LoRaWAN antenna. Let $G_g(n)$ denote the path gain in decibels from gateway $g \in \{1,\ldots,G\}$ to grid cell $n \in \{1,\ldots,N\}$. This term captures free-space loss, building interactions, and material-dependent attenuation along the propagation paths. The LoRaWAN configuration follows the RAK7289 WisGate Edge Pro outdoor gateway from RAKwireless and a typical Class~A end device. The main parameters used in the simulations are summarized in Table~\ref{tab:lora_params}.

\begin{table}[t]
    \caption{LoRaWAN link budget parameters used in the simulations.}
    \label{tab:lora_params}
    \centering
    \begin{tabular}{l l l}
        \hline
        Parameter & Symbol & Value \\
        \hline
        Carrier frequency & $f_c$ & \SI{1.0}{\giga\hertz}\footnotemark \\
        System bandwidth & $B$ & \SI{125}{\kilo\hertz} \\
        Gateway transmit power & $P_\text{tx}$ & \SI{20}{dBm} \\
        Receiver noise figure & $F_\text{NF}$ & \SI{6}{\dB} \\
        Ambient temperature & $T_\text{amb}$ & \SI{290}{\kelvin} \\
        Implementation margin & $M$ & \SI{10}{\dB} \\
        Robust SNR threshold & $\gamma_\text{rob}$ & \SI{-10}{\dB} \\
        Edge SNR threshold & $\gamma_\text{edge}$ & \SI{-22}{\dB} \\
        \hline
    \end{tabular}
\end{table}
\footnotetext{The ITU material models in Sionna are defined above \SI{1}{\giga\hertz}. A carrier of \SI{1.0}{\giga\hertz} is therefore used as a proxy for the \SIrange{915}{923}{\mega\hertz} LoRa band; the resulting difference in free-space loss over the distances considered is below \SI{1}{\dB}.}

The thermal noise power including receiver noise figure is
\begin{equation}
    N_\text{dBm} = 10\log_{10}\!\left(\frac{k T_\text{amb} B}{1~\text{mW}}\right) + F_\text{NF},
\end{equation}
\noindent where $k$ is Boltzmann’s constant. The per gateway SNR at grid cell $n$ is then defined as:
\begin{equation}
    \text{SNR}_g(n) = P_\text{tx} + G_g(n) - N_\text{dBm} - M \quad [\text{dB}],
    \label{eq:snr_def}
\end{equation}
\noindent where $M$ represents implementation margins that are not explicitly modeled in the ray tracing engine.

Coverage is defined with respect to a single robust SNR threshold $\gamma_\text{thr} = -10~\text{dB}$, which corresponds to a  decodable uplink for moderate spreading factors. For each gateway $g$ and cell $n$, we form a binary coverage indicator
\begin{equation}
    c_g(n) = \mathbb{I}\!\left\{\text{SNR}_g(n) \geq \gamma_\text{thr}\right\}.
\end{equation}

Next, regarding the gateway selection metric, let $\mathcal{G}=\{1,\ldots,G\}$ denote the set of gateways and $\mathcal{C}=\{1,\ldots,N\}$ the set of grid cells. For a threshold $\gamma$, the coverage set of gateway $g$ is
\begin{equation}
    \mathcal{C}_g = \bigl\{ n \in \mathcal{C} : \text{SNR}_g(n) \geq \gamma_\text{thr} \bigr\}.
\end{equation}
For any subset of gateways $\mathcal{S} \subseteq \mathcal{G}$, the union coverage and the corresponding area fraction are
\begin{align}
    \mathcal{C}(\mathcal{S}) &= \bigcup_{g \in \mathcal{S}} \mathcal{C}_g, \\
    f(\mathcal{S}) &= \frac{|\mathcal{C}(\mathcal{S})|}{|\mathcal{C}|}.
\end{align}

The objective is to maximize $f(\mathcal{S},\gamma)$ subject to a budget $|\mathcal{S}|\leq K$. This is the standard budgeted maximum coverage problem, in which the coverage function is submodular, so a greedy algorithm that iteratively adds the gateway with the largest increase in $|\mathcal{C}(\mathcal{S},\gamma)|$ provides a better $(1-1/e)$ approximation and a clear ranking of candidate rooftops.

\section{Results and Discussion}
\label{sec:results}

\begin{figure}[t]
  \centering
  \includegraphics[width=\linewidth]{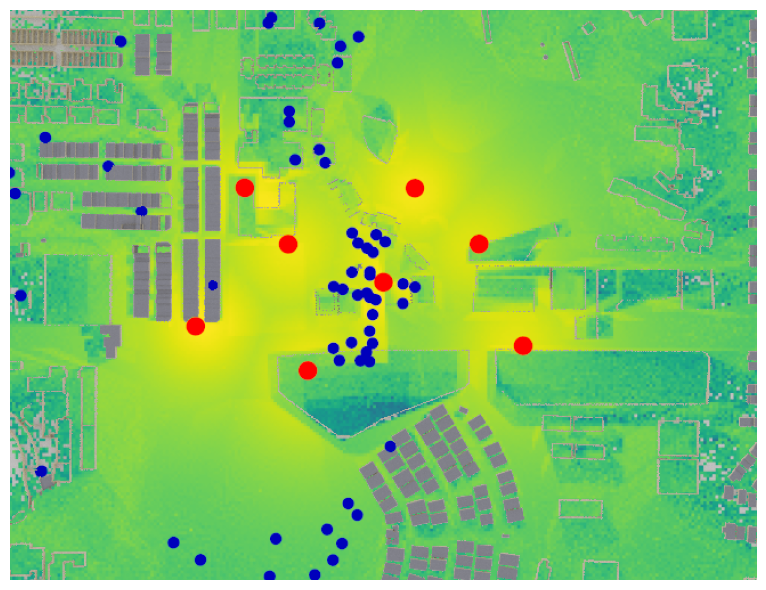}
  \caption{LoRaWAN path gain map over the DT. Red markers denote candidate rooftop gateways and blue markers denote sensor locations.}
  \label{fig:radiomap_sensors}
\end{figure}

Figure~\ref{fig:radiomap_sensors} presents a view of the DT (i.e., Sunway City replica). The Figure shows the locations of the eight gateways represented with red markers. We have added 100 sensors randomly to represent a dense deployment of IoT. The resulting map shows good signal along the open streets and near the water area. It also shows shadow regions near the tallest building and inner courtyards of the campus. This pattern shows that the ray tracing engine is capturing the expected propagation behavior of a sub-GHz rooftop deployment, and highlights locations where a high sensitivity LoRa link would be challenging.

\begin{figure}[t]
  \centering
  \includegraphics[width=\linewidth]{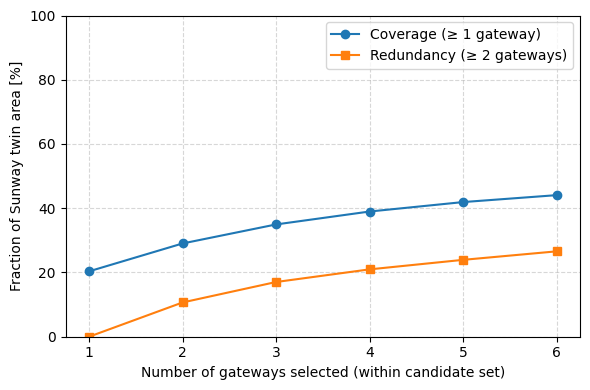}
  \caption{Fraction of the DT that is covered by at least one gateway (i.e., blue) and by at least two gateways (i.e., orange) as a function of the number of deployed gateways selected from the candidate rooftops. Coverage is defined with respect to the robust SNR threshold of $-10$~dB.}
  \label{fig:coverage_vs_K}
\end{figure}
Figure~\ref{fig:coverage_vs_K} and Table~\ref{tab:coverage_vs_K} summarize how coverage over the DT changes as gateways are added according to the greedy selection. For the robust SNR threshold of $-10$~dB, a single gateway covers $20.35\%$ of the grid cells ($K=1$). With two and three gateways the coverage fraction increases to $29.03\%$ and $34.96\%$, respectively, and reaches $44.07\%$ with the selected six gateways. The growth is sub-linear, in which, the third gateway adds $5.93$ percentage points of additional area, whereas the sixth adds only $2.15$ percentage points. This behavior is consistent with the set cover structure of the problem; meaning that once the main street area and open spaces are covered, additional rooftops fill in smaller areas of shadow.

Figure~\ref{fig:coverage_vs_K} also shows the fraction of the area that has redundant connectivity, defined as coverage by at least two distinct gateways above the robust SNR threshold. For $K=1$ this fraction is zero, because a single gateway cannot provide diversity. When we add a second gateway, $10.66\%$ of the city has two simultaneous links. This fraction increases to $17.03\%$ at $K=3$ and $26.58\%$ at $K=6$. In other words, when we deploy the six gateways, $44.07\%$ of the DT is covered by at least one gateway and $26.58\%$ is covered by two or more gateways. The gap between the two curves in Figure~\ref{fig:coverage_vs_K} quantifies the area that is served by a single robust link. However, this lacks macro-diversity and it is therefore more sensitive to gateway outages or shadowing.

\begin{table}[t]
    \centering
    \caption{Coverage and redundancy over the DT.}
    \label{tab:coverage_vs_K}
    \begin{tabular}{ccc}
        \hline
        $K$ & Coverage $\ge 1$ GW [\%] & Coverage $\ge 2$ GWs [\%] \\
        \hline
        1 & 20.35 & 0.00 \\
        2 & 29.03 & 10.66 \\
        3 & 34.96 & 17.03 \\
        4 & 38.98 & 20.97 \\
        5 & 41.92 & 23.93 \\
        6 & 44.07 & 26.58 \\
        \hline
    \end{tabular}
\end{table}

\begin{figure}[t]
  \centering
  \includegraphics[width=\linewidth]{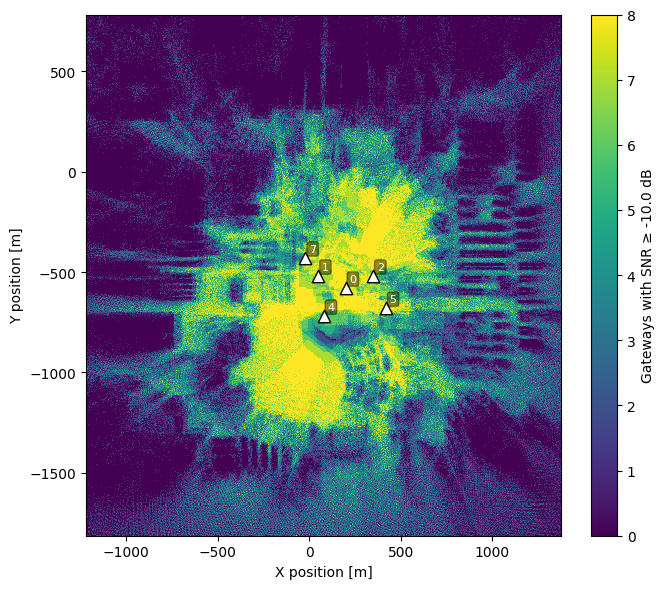}
  \caption{Number of gateways that cover each grid cell at the robust SNR threshold. White triangles indicate candidate rooftop locations.}
  \label{fig:redundancy}
\end{figure}

Coverage redundancy is shown in Figure~\ref{fig:redundancy}, which plots, for each grid cell, the number of gateways satisfying the robust SNR threshold. The central campus region exhibits three to six overlapping gateways, suggesting that diversity combining or simple gateway selection can significantly improve reliability in that zone. In contrast, other areas towards the edges of the DT (e.g., residential areas) have zero or one covering gateway. These areas can bee greatly affected by additional attenuation from foliage, vehicles or indoor placement. From a network planning perspective, this map shows which areas of the city would benefit most from either additional gateways or higher mounting points if the deployment is expected to support sensors beyond that specific area (i.e., in this case the university campus).

\begin{figure}[t]
  \centering
  \includegraphics[width=\linewidth]{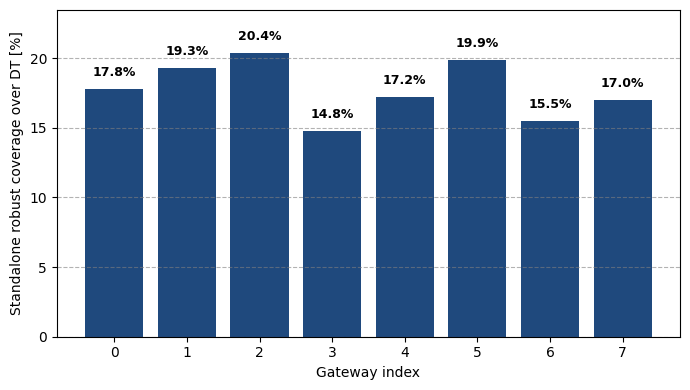}
  \caption{Standalone robust coverage fraction when each candidate rooftop is used alone as a LoRaWAN gateway.}
  \label{fig:standalone}
\end{figure}

Figure~\ref{fig:standalone} shows the standalone robust coverage when each candidate rooftop is used individually as a LoRaWAN gateway. The x-axis indexes rooftops by decreasing standalone coverage; the number above each bar indicates the original gateway index in the candidate set.  Gateway~2 provides the largest standalone footprint of the grid cells. Gateways~3 and~6 cover \(14.8\%\) and \(15.5\%\) of the area, indicating that these two rooftops are weaker candidates when used as individual deployment.

The greedy selection that maximizes union coverage (i.e., the total number of distinct grid cells covered when gateways are combined) for budgets \(K=1,\dots,6\), and chooses gateways in this order \([2,1,5,0,7,4]\). This sequence is consistent with the standalone findings, in which, the first three greedy choices are the rooftops with the three largest individual coverage, and the remaining selected rooftops (Gateways 0, 7, 4) are in the middle of the standalone distribution. The two weakest rooftops in Figure~\ref{fig:standalone} (Gateways 6 and 3) are never selected by the greedy algorithm within the considered budget, since they contribute little unique area other than what is already covered by the stronger sites. This alignment between the standalone statistics and the greedy outcome supports the interpretation that height and centrality dominate the placement decision.

\begin{figure}[t]
  \centering
  \includegraphics[width=\linewidth]{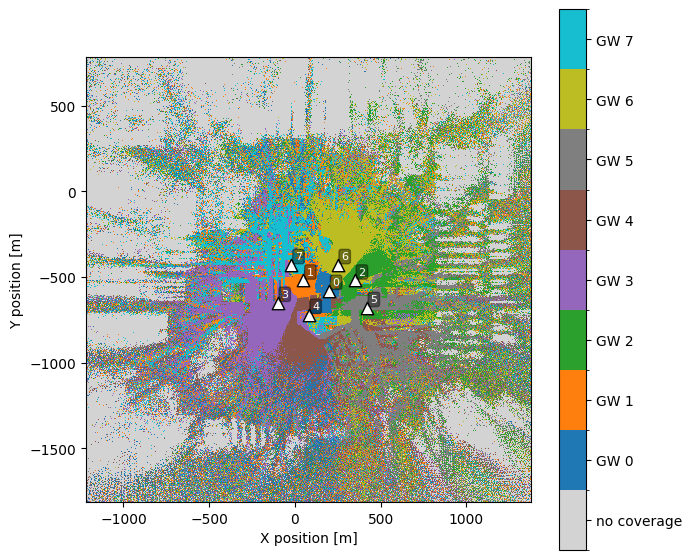}
  \caption{Best gateway association map at the robust SNR threshold. Each color represents the gateway providing the highest SNR; grey cells are below the threshold. White triangles denote gateway locations.}
  \label{fig:association}
\end{figure}

The best gateway association map in Figure~\ref{fig:association} assigns each grid cell to the gateway that provides the highest robust SNR, with grey cells indicating locations below \(\gamma_\text{thr}\). The resulting service regions are irregular and they do not follow simple circular or Voronoi-like shapes. Instead they follow street orientations, open spaces and gaps between buildings. Some gateways have disjoint association regions connected by propagation corridors, which would not be visible under the conventional distance-based model. For network planning, this finding is useful for dimensioning the backhaul capacity of each gateway, and for anticipating, where gateway handovers or roaming between operators are would occur.

\section{Conclusion}
\label{sec:conclusion}

This paper used a geometry accurate DT of Sunway City together with Sionna’s ray-tracing engine to examine LoRaWAN gateway placement under a commercial gateway link budget. By computing path gains on a \SI{5}{\meter} grid on the DT and translating them into SNR based coverage indicators, the study quantified how coverage and redundancy evolve as gateways are added according to a greedy maximum coverage criterion. The results show that a single rooftop gateway not sufficient for the selected dense area (i.e., the university campus). Even with six carefully selected sites, the robust SNR threshold is met in only \(44.07\%\) of the DT, and macro-diversity with at least two gateways is possible in  \(26.58\%\) of the area. Redundancy maps show that diversity is concentrated over the campus. Standalone coverage findings and best gateway association maps further show that some rooftops are dominated, whereas a few central rooftops support extended propagation corridors that would have been not possible to anticipate using distance based planning. Future studies can extend this work, by adding empirical based spreading factor and traffic distributions. Future studies can also explore the implications for different antenna heights or mast structures, and formulate multi-objective placement criteria that trade off coverage, redundancy, and backhaul constraints. The DT provides a general approach, in which frameworks or models developed using the DT can be applied to other similar dense cities.

\section*{Acknowledgment}

The authors thank and gratefully acknowledge the support of Sunway University for funding this work under the Research Accelerator Seed Grant, Ref No: GRTIN-RAG(02)-DEN-03-2024.


\bibliographystyle{ieeetr}
\bibliography{IEEEexample}


\vspace{12pt}

\end{document}